\documentclass[prb,twocolumn,showpacs,superscriptaddress,preprintnumbers,amsmath,amssymb]{revtex4}

\usepackage{graphicx}
\usepackage{color}
\usepackage[colorlinks=true,plainpages=false,linkcolor=blue,urlcolor=blue,citecolor=blue,pdfpagemode=UseNone,pdfstartview=FitBH]{hyperref}

\def\HG#1 {\emph{\color{blue}#1}}

\begin{document}

\title{Light-Induced Metastable State in Charge-Ordered YBa$_2$Cu$_3$O$_{6+x}$}

\author{H.~Gretarsson}
\author{S. M. Souliou}
\author{S.~Jeong}
\author{J.~Porras}
\author{T.~Loew}
\affiliation{Max-Planck-Institut f\"{u}r Festk\"{o}rperforschung, Heisenbergstr. 1, D-70569 Stuttgart, Germany}
\author{M.~Bluschke}
\affiliation{Max-Planck-Institut f\"{u}r Festk\"{o}rperforschung, Heisenbergstr. 1, D-70569 Stuttgart, Germany}
\affiliation{Helmholtz-Zentrum Berlin fur Materialien und Energie, BESSY II, Albert-Einstein-Str. 15, 12489 Berlin, Germany}
\author{M.~Minola}
\author{B. Keimer}
\email[]{b.keimer@fkf.mpg.de}
\affiliation{Max-Planck-Institut f\"{u}r Festk\"{o}rperforschung, Heisenbergstr. 1, D-70569 Stuttgart, Germany}
\author{M. Le Tacon}
\email[]{matthieu.letacon@kit.edu}
\affiliation{Max-Planck-Institut f\"{u}r Festk\"{o}rperforschung, Heisenbergstr. 1, D-70569 Stuttgart, Germany}
\affiliation{Karlsruher Institut f\"{u}r Technologie, Institut f\"{u}r Festk\"{o}rperphysik, Hermann-v.-Helmholtz-Platz 1, D-76344 Eggenstein-Leopoldshafen, Germany}

\date{\today}

\begin{abstract}

We report temporal changes in Raman-scattering spectra of detwinned YBa$_2$Cu$_3$O$_{6+x}$ single crystals under exposure to red laser light polarized along the crystallographic $a$-axis. A recent publication by Bakr \textit{et al.}  (\href{http://link.aps.org/doi/10.1103/PhysRevB.88.214517}{Phys. Rev. B {\bf 88}}, \href{http://link.aps.org/doi/10.1103/PhysRevB.88.214517}{214517} (\href{http://link.aps.org/doi/10.1103/PhysRevB.88.214517}{2013})) identified new Raman-active modes that appear upon cooling below $T \sim 200$ K, and attributed these modes to charge ordering phenomena observed in x-ray scattering and nuclear magnetic resonance experiments on the same materials. Here we report that the intensities of these modes depend not only on temperature and oxygen content, but also on the cumulative photon dose absorbed by the YBa$_2$Cu$_3$O$_{6+x}$ samples. The light-induced changes in the Raman spectra exhibit a stretched-exponential time dependence, with a characteristic time that increases strongly upon cooling. They also depend strongly on doping and are most pronounced for $x \sim 0.6$. In fully illuminated samples, the mode intensities exhibit a monotonic temperature dependence indicative of a second-order phase transition. The findings indicate a metastable state generated by light-induced rearrangement of the oxygen dopants. We hypothesize that the new Raman phonons are associated with a three-dimensional charge-ordered state induced  by light illumination, analogous to a state that was recently observed by resonant x-ray scattering in oxygen-disordered YBa$_2$Cu$_3$O$_{6+x}$ films.

\end{abstract}


\maketitle

\noindent

\section{Introduction}

The YBa$_2$Cu$_3$O$_{6+x}$ family of high-temperature superconductors (hereafter YBCO$_{6+x}$) has long served as a model system for microscopic theories of unconventional superconductivity. One of the hallmarks that sets this system apart from related compounds, such as the Bi- and Hg-based cuprates, is the presence of copper oxide chains that run along the crystallographic $b$-axis and act as charge reservoirs for the CuO$_2$ planes. When the chains are completely depleted of oxygen ($x=0$),  there is one hole per Cu site in the CuO$_2$ planes, and strong electronic correlations induce a charge-transfer insulating state. Adding O(1) ions into interstitial sites leads to the formation of Cu-O(1) chain fragments, and eventually to a transfer of mobile holes to the CuO$_2$ planes. \cite{Zaanen_PRL1988} Superconductivity appears for $x \gtrsim 0.3$, and the superconducting transition temperature $T_c$ increases with $x$ in the ``underdoped'' regime,  that is, until $T_c$ reaches its optimal value of $\sim 93$ K for $x \sim 0.93$.

The high mobility of the O(1) ions gives rise to non-equilibrium phenomena with a profound influence on the transport properties. For instance, room temperature annealing of chain-disordered YBCO$_{6+x}$ crystals for several hours can enhance the Cu-O(1) chain order and increase $T_c$~\cite{Veal_PRB1990, annealing}. The dependence of $T_c$ on annealing time decreases as the diffusion of the O(1) atoms is progressively quenched with decreasing temperature, and it vanishes entirely below $T \sim 250$ K. At lower temperatures, the O(1) mobility can be restored by exposing YBCO$_{6+x}$ to visible light, which results in increased electrical conductivity and enhanced $T_{c}$ \cite{Kudinov_JETP1990,Kudinov_PhysLettA1990,Suzuki_PRL1994}, analogous to the effect of room temperature annealing. These observations indicate that visible light allows the otherwise immobile O(1) atoms to rearrange themselves into longer CuO chains that transfer more mobile holes to the CuO$_2$ layers \cite{Markowitsch_PhysC1996,Kudinov_PRB1993}. This scenario is supported by Raman scattering experiments that revealed a gradual light-induced ``bleaching'' of Raman-forbidden modes originating from copper and oxygen vibrations at the end of short CuO chain fragments~\cite{Wake_PRL1991,Ivanov_PRB1995,Bahrs_PRB2004}. Both the photoconductivity and the bleaching of the Raman modes persist when the laser beam is switched off at low temperatures. Heating the system above $T\approx 250$ K restores the intensity of the defect modes and erases the persistent photoconductivity  \cite{Kall_PRB1998}, indicating that the light-induced state (hereafter LIS$_1$) is metastable. 

Appropriate heat treatments promote the formation of Cu-O(1) chains with lengths exceeding 10 nm~\cite{Zimmermann_PRB2003}. The low level of disorder, compared to other superconducting cuprate families with randomly placed dopants, has enabled the observation of quantum oscillations~\cite{Doiron_Nature2007} in high magnetic fields over a wide range of doping levels, even in the underdoped regime. High-field transport experiments indicate a Fermi surface reconstruction by a long-range electronic superstructure for underdoped samples with 0.45 $\lesssim$ x $\lesssim$ 0.86~\cite{Sebastian_PNAS2010, Ramshaw_Science2015}. The quasi-two-dimensional charge density wave (CDW) recently discovered in this material\cite{Ghiringhelli_Science2012,Achkar_PRL2012,Chang_NaturePhysics2012,Wu_Nature2011} in the same doping range has been proposed as the origin of the Fermi surface reconstruction. The observation of CDWs in several other compound families demonstrates that this phenomenon is generic for the layered cuprates~\cite{Comin_Science2014,Tabis_NatCom2014,daSilvaNeto_Science2014,daSilvaNeto_Science2015}.

Although the relationship between oxygen order and CDW formation in YBCO$_{6+x}$ has been addressed in prior work, a conclusive picture has not yet emerged.  Some experiments indicate a link between both phenomena. For instance, x-ray scattering experiments have shown that the CDW order becomes highly anisotropic in the presence of long-range oxygen order, including the ``ortho-II'' structure where full and empty oxygen chains alternate ~\cite{Blanco_PRL2013,Blanco_PRB2014,Huecker_PRB2014,Blackburn_PRL2013}. On the other hand, neither the wave vector nor the correlation length of the CDW are significantly affected by annealing-induced modifications of the oxygen order (although some infuence on the CDW amplitude has been observed) \cite{Achkar_PRL2014}. These results call for further investigations of the interplay between the CuO chain order and the CDW. 

Here we investigate the photon illumination dependence of a set of phonon modes that were recently observed in Raman scattering experiments on underdoped YBCO$_{6+x}$ single crystals upon cooling below $\sim 200$ K. Since the onset temperatures of the new phonons is close to the one of the CDW, the phonons were attributed to folding of the Brillouin zone that is expected in the CDW state ~\cite{Bakr_PRB2013}. The modes have the $A_g$ symmetry and are strongly resonant with red incident light. We report a strong illumination dependence of these modes in light polarized along the crystallographic $a$-axis, which we attribute to a second light-induced state (LIS$_2$) coexisting with the CDW. The dependence of LIS$_2$ on doping and temperature suggests that it originates from a reordering of the O(1) atoms in the CuO chains. Although the LIS$_2$ shares many properties with the aforementioned LIS$_1$ induced by $b$-axis polarized light \cite{Wake_PRL1991}, their temperature dependences, doping dependences, and phonon spectra are distinct. Based on a comparison with recent resonant x-ray scattering results on oxygen-disordered YBCO$_{6+x}$ films, \cite{Bluschke_NComm2018} we hypothesize that the phonon modes in the LIS$_2$ arise from Brillouin-zone folding in a three-dimensionally charge-ordered state.

\section{Experimental Details}

\subsection{Samples}

High-quality detwinned and freshly cleaved YBCO$_{6+x}$ single crystals were used for the measurement. Details of the growth and characterization of the crystals have been reported in earlier publications.\cite{Lin_JCG2002,Blanco_PRB2014} The hole doping level $p$ was determined by measuring  
$T_c$ and the out-of-plane lattice parameter $c$ of each sample as described in Ref. \onlinecite{Liang_PRB2006}.

\subsection{Raman measurements}

The Raman measurements were performed in backscattering geometry using a Jobin Yvon LabRam 1800 single grating spectrometer equipped with a razor-edge filter. The  $\lambda = 633$ nm (1.96 eV) line of a HeNe laser was used for the $XX$ channel, and the $\lambda = 532$ nm (2.33 eV) line of a diode laser for the $YY$ channel. The laser power was $\sim 1$ mW, and the beam diameter was $\sim 10~\mu$m. In these exact experimental conditions local heating effects  have been found to be less than 5 K (see Ref. \onlinecite{Bakr_PRB2013}). The samples were cooled in a He-flow cryostat. Each spectrum was corrected for the Bose factor. To simplify the following discussion, we use the terms ``un-illuminated'' for spectra from samples after short (1 min.) exposure to visible light, and ``illuminated'' for long ($> 30$ min.) exposure. We consider the samples ``fully illuminated'' when the spectrum shows no observable change after 30 minutes. The same laser line is used to illuminate the samples and to probe the Raman excitations, guaranteeing identical sample volumes for illumination and data collection.

To describe the scattering geometries, we use the Porto notation A(BC)A$^\prime$, where A(B) and A$^\prime$(C) stand for the propagation (polarization) directions of the incident and scattered light, respectively, relative to the crystalline axes. All of the data presented in this paper were taken with the light propagation direction along the crystalline $c-$axis; hence only the polarization orientations will be specified.

\begin{figure}[htb]
\includegraphics[width=\columnwidth]{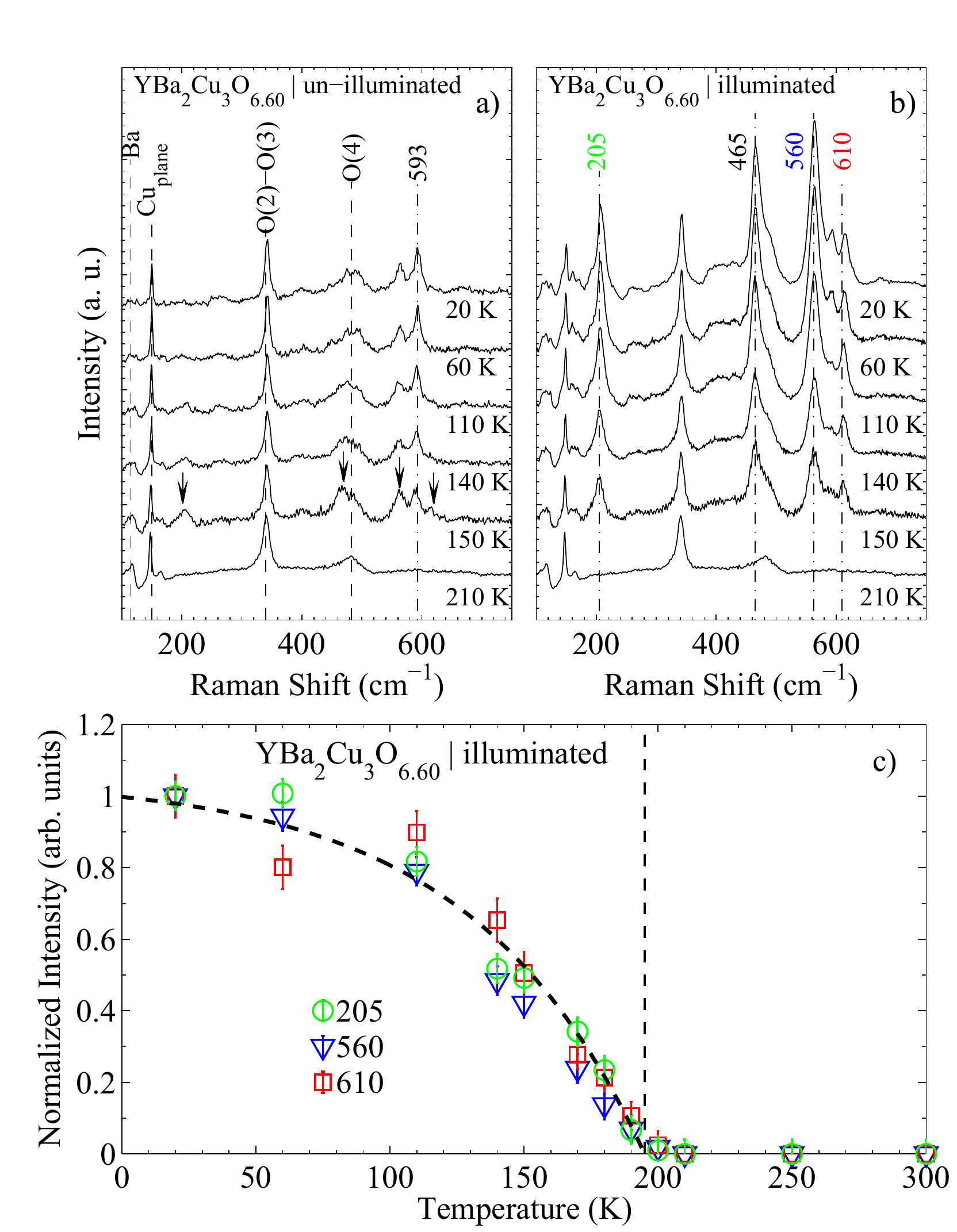}
\caption{\label{fig01}(Color online) Temperature dependence of the Raman spectra for YBCO$_{6.6}$ taken in the $XX$ channel using a red laser line ($\lambda=633$ nm). Two different laser exposures were used (see text): (a) un-illuminated and (b) illuminated. For each scan a spot on the sample that had not been exposed to the laser was used. The arrows in panel (a) mark the phonons that are most susceptible to the laser exposure. (c) Temperature dependence of the integrated intensity of the modes at 205, 560, and 610-cm$^{-1}$.}
\end{figure}

\section{Experimental Results}

\subsection{Illumination effect}

Ref.~\onlinecite{Bakr_PRB2013} demonstrated that new modes appear upon cooling below $\sim 200$ K in the Raman spectra of underdoped YBCO$_{6+x}$ samples that also show a CDW signal in x-ray scattering experiments~\cite{Blanco_PRB2014}. Further investigation to unveil the nature of these new modes revealed that, at low temperature, their intensities are strongly dependent on the laser exposure time. To illustrate this illumination effect, we first present the Raman spectra of YBCO$_{6.6}$ ($T_c=61$ K) in the $XX$ channel. Fig. \ref{fig01} (a) shows the temperature dependence of un-illuminated Raman spectra using an incident laser wavelength of $\lambda=633$ nm. In order to ensure that the sample was kept un-illuminated, the laser beam was switched off during cooling and a new spot was used for each temperature. Above $T = 210$ K the spectra consist of a broad, featureless electronic continuum with superimposed sharp optical phonons. The observed four Raman-active modes can be associated with the fully oxygenated ($x=1$) ``ortho-I'' unit cell\cite{Liu_PRB1988,Iliev_PRB2008}: 1) Ba($z$) at $\sim$115 cm$^{-1}$, 2) Ba/Cu(2)($z$) at $\sim$150 cm$^{-1}$, 3) O(2)-O(3)($z$) at $\sim$340 cm$^{-1}$, and 4) O(4)($z$) at $\sim$485 cm$^{-1}$.  Upon cooling the electronic continuum is suppressed along with the appearance of additional Raman-active modes at $\sim$205 cm$^{-1}$, $\sim$465 cm$^{-1}$, $\sim$560 cm$^{-1}$, $\sim$593 cm$^{-1}$ ,  and $\sim$610 cm$^{-1}$. These modes, marked by arrows in Fig. \ref{fig01} (a), have the strongest intensity at $T=150$ K and are considerably weaker or absent at $T=20$ K.

In Fig. \ref{fig01} (b) we plot the illuminated spectra of YBCO$_{6.6}$ as a function of temperature. In striking contrast to the data in the un-illuminated state, the phonon modes marked with arrows in (a) become much stronger and do not exhibit any intensity reduction at low temperatures. We note that this behavior is not seen for the $\sim$593 cm$^{-1}$ mode which is related to the Raman forbidden O(1) vibration within broken CuO chains~\cite{Liu_PRB1988}. The new modes are long-lived at low temperatures, as no intensity reduction was observed when the laser was turned off for 12 hours at $T = 20$ K. In Fig. \ref{fig01} (c) the temperature dependence of the integrated intensity of the fully illuminated $\sim$205 cm$^{-1}$,  $\sim$560 cm$^{-1}$, and $\sim$610 cm$^{-1}$ modes can be seen. The modes disappear above $T = 200$ K, as reported in Ref.~\onlinecite{Bakr_PRB2013},  but no suppression of the intensity is found below $T_{c} = 61$ K (see the discussion in Section IV). This indicates that the new modes originate from a metastable state (termed LIS$_2$), which is reminiscent of the well studied LIS$_1$ ``bleached'' state\cite{Wake_PRL1991,Ivanov_PRB1995,Bahrs_PRB2004} that arises from the light-induced oxygen reordering in the CuO chains. In order to check the relationship of the newly found LIS$_2$ to the LIS$_1$ and to the CDW, we investigated the dependence of the intensity of the new modes on the illumination time, as well as the hole doping dependence of the illumination effect.

\begin{figure}[htb]
\includegraphics[width=\columnwidth]{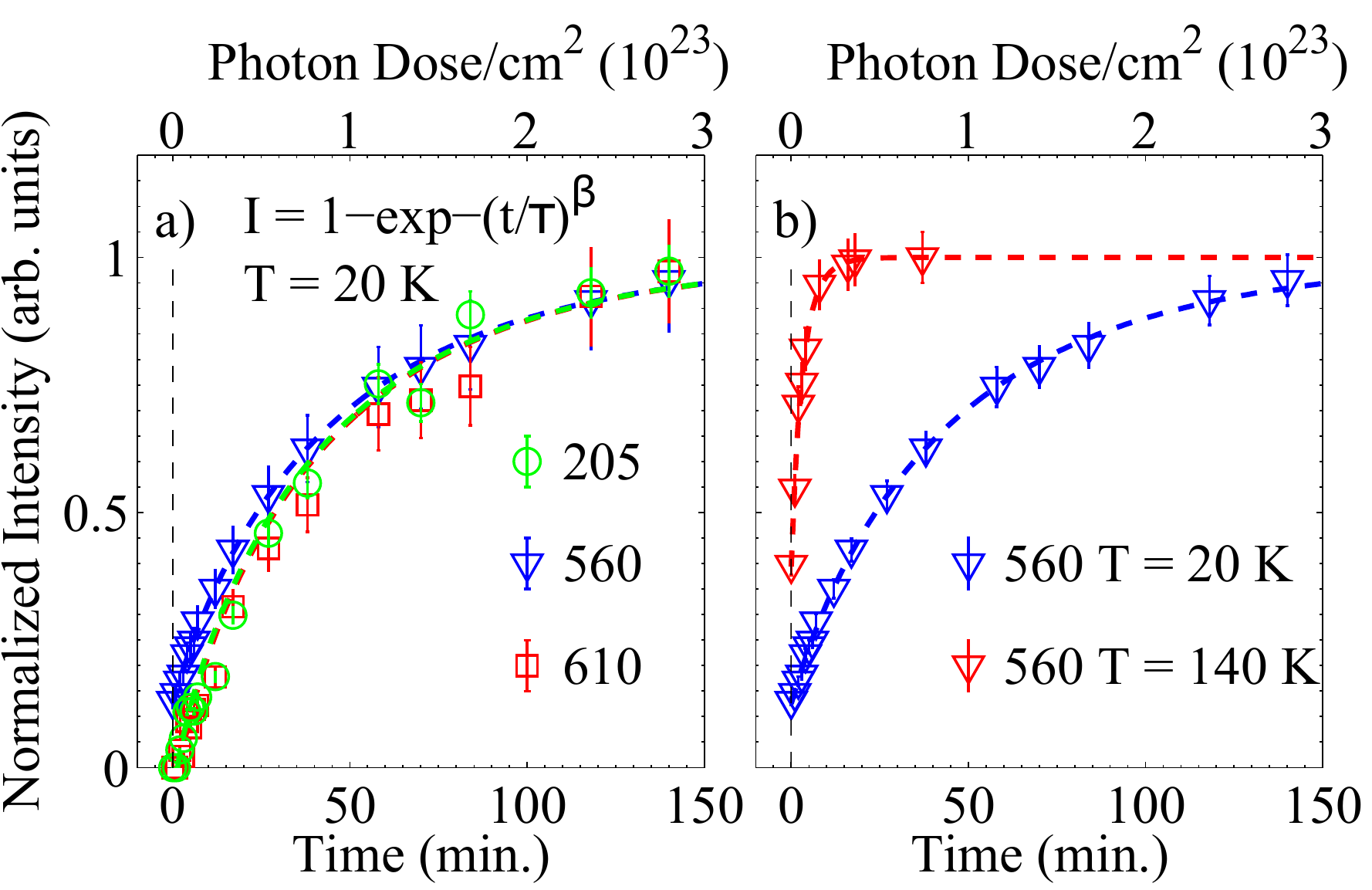}
\caption{\label{fig02}(Color online) (a) Time dependence of the integrated intensity of selected light-induced phonon modes at $T = 20$ K. Top axis shows the cumulative photon dose. Dashed lines are the results of fits to a stretched exponential function (see text). (b) Comparison of the time dependence of the integrated intensity of the 560 cm$^{-1}$ phonon mode for $T = 20$ K and $T = 140$ K.   }
\end{figure}

\subsection{Photon dose dependence}

To quantify the illumination dependence of the new modes in LIS$_2$, we plot in Fig. \ref{fig02} (a) the integrated intensity of the $\sim$205 cm$^{-1}$, $\sim$560 cm$^{-1}$, and $\sim$610 cm$^{-1}$ phonons at $T = 20$ K as a function of time (bottom axis) or cumulative photon dose per unit area (top axis). Both quantities are proportional as the beam power was kept constant. The integrated intensity was extracted by fitting each phonon mode with a Lorentzian function using a fixed width and peak position. Spectra at different times were recorded on the same spot with one minute acquisition each.

A good description of the time ($t$) dependence of the integrated intensity is obtained with a stretched exponential function  $I(t)=1-exp(\frac{-t}{\tau})^{\beta}$, were $\tau$ represents the average characteristic time scale and $\beta = 0.85$ is the stretching exponent. Stretched exponentials with $\beta<1$ are commonly used to describe disordered systems with a broad range of characteristic times, and this type of relaxation is often found in glassy materials~\cite{Welch_PRL2013}. The strong 560 cm$^{-1}$ mode can already be seen in the un-illuminated spectrum (Fig. \ref{fig01} (a)), while the 205 cm$^{-1}$ and 610 cm$^{-1}$ modes are suppressed. This causes a small discrepancy in the intensity evolution of these phonons at small photon dose in Fig.  \ref{fig02} (a), but the overall trend is comparable for all of the new modes.

In Fig. \ref{fig02} (b) we compare the illumination dependence of the 560 cm$^{-1}$ mode at two temperatures, $T = 20$ K and $T = 140$ K. The characteristic time scale for the intensity increase of the mode is strongly reduced as temperature is increased. By using the same fitting function as at low temperature (keeping $\beta = 0.85$) we obtain  $\tau = 40$ minutes and $4$ minutes at $T= 20$ K and 140 K, respectively. We have checked several temperatures and found a smooth decrease of $\tau$ with increasing temperature, without any noticeable anomaly at $T_c$.

\subsection{Hole doping dependence}

\begin{figure}[htb]
\includegraphics[width=1\columnwidth]{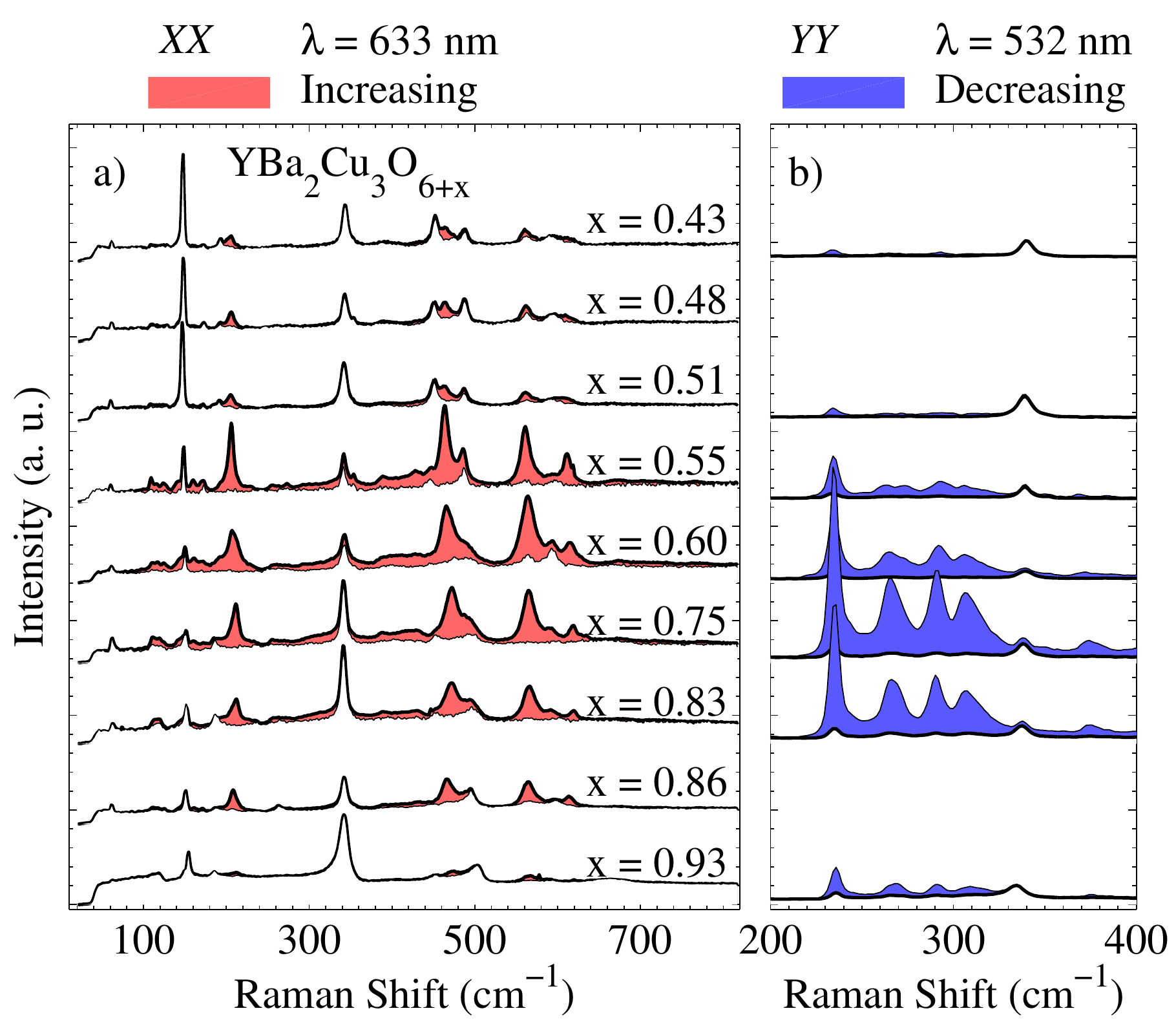}
\caption{\label{fig03} Doping dependence of the new (a) $XX$ light-induced phonon modes in the LIS$_2$ and (b) $YY$ light-induced suppression of CuO chain defect phonon modes  in the LIS$_1$ taken at $T = 20$ K. For each hole doping a spectrum was taken on an un-illuminated fresh spot and compared with a spectrum taken with full illumination. The highlighted red area in (a) indicates the light-induced enhancement, the highlighted blue area in (b) the light-induced suppression. Note that in (b) a green laser line was used in order to be closer to the CuO chain defect resonance.}
\end{figure}

\begin{figure}[htb]
\includegraphics[width=\columnwidth]{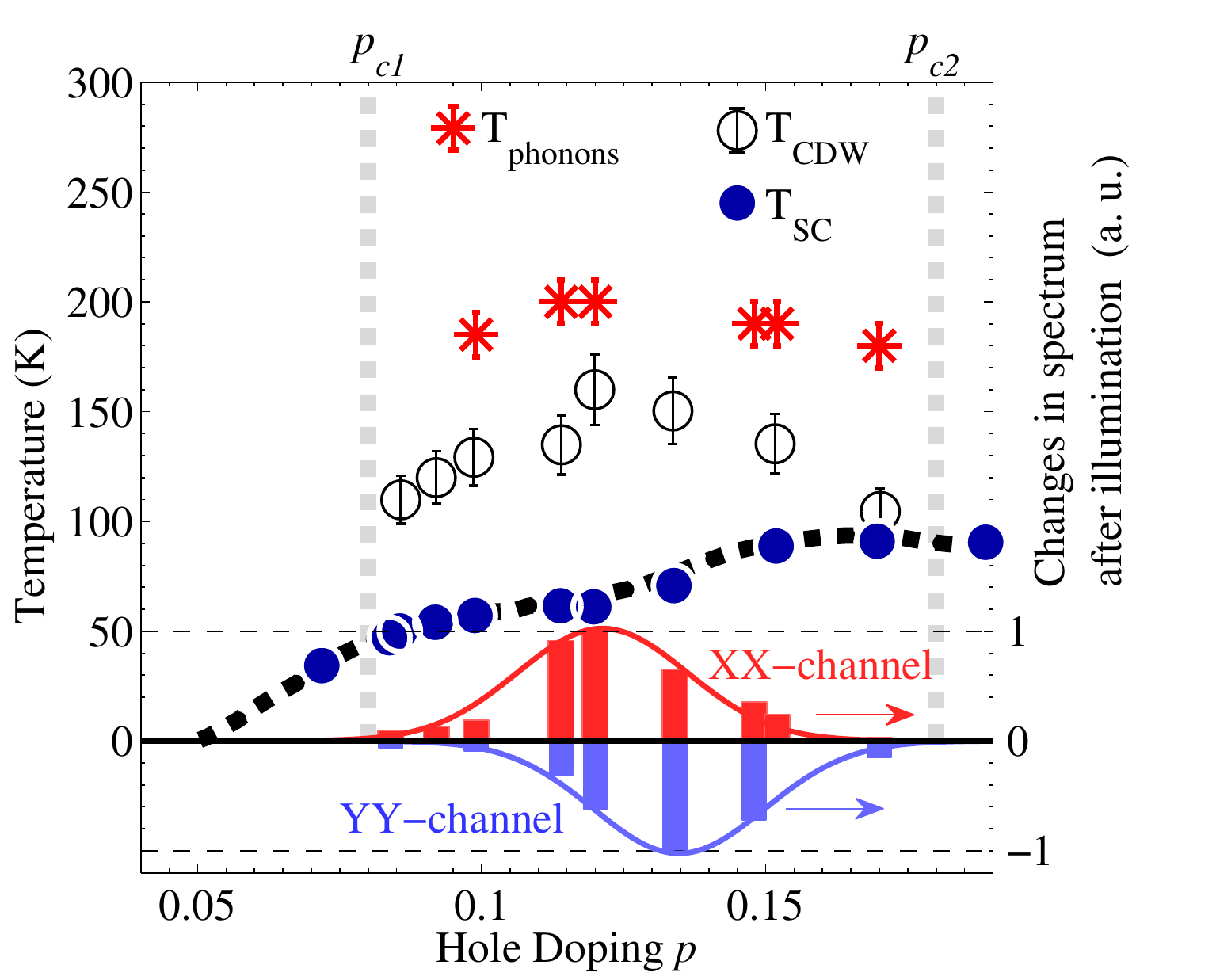}
\caption{\label{fig04}(Color online) Doping dependence of the strength of the LIS$_2$ (red bars) and the LIS$_1$ (blue bars). Selected onset temperatures (red stars) for the new phonon modes in the LIS$_2$ are also presented. The LIS$_1$ is already observed at T = 300 K. The data are overlaid with results from Ref. \onlinecite{Blanco_PRB2014}. Closed blue circles represent the measured $T_{c}$, the open circles represent the onset of the CDW. The dashed vertical lines, represented by $p_{c1}$ and  $p_{c2}$, mark the hole-doping range where CDW is observed by x-ray scattering. }
\end{figure}

Next, we discuss the hole doping dependence of LIS$_2$. In Fig. \ref{fig03} (a) we plot the un-illuminated Raman spectra (thin line) of YBCO$_{6+x}$ for selected oxygen concentrations. Each spectrum was taken at $T = 20$ K in the $XX$ channel. Because of the very large renormalization of the 340 cm$^{-1}$ phonon mode in the $x=0.93$ sample across $T_{c}$, we only show the spectrum taken at $T = 90$ K. This facilitates the comparison with the other doping levels. For all samples we observe the same four new phonon modes after full illumination (thick line, changes are highlighted in red), albeit with different intensities. In Fig. \ref{fig04} we plot the ratio of the integrated spectrum (from 50-800 cm$^{-1}$)  for the illuminated and un-illuminated cases (red bars) as a function of hole-doping $p$ (here we have normalized the largest change to unity).  These measurements show that the LIS$_2$ is strongest at $x\!\approx 0.60$ ($p\!\approx 0.12$) and is quickly suppressed at both lower and higher doping levels.  We have also measured the onset temperature of the new modes, T$_{\rm phonons}$, for selected samples, and find that T$_{\rm phonons} = 190$ K for $x = 0.51$, 200 K for $x = 0.6$ where LIS$_2$ is strongest, and 180 K for $x = 0.93$.

\subsection{Comparison with the ``bleached'' state LIS$_1$}

Insight into the origin of the LIS$_2$ can be gained by comparing our measurements with the better understood ``bleached'' state (LIS$_1$). We address this issue by carrying out the same doping dependent measurements as in Fig. \ref{fig03} (a), but using the $YY$ channel and green laser light ($\lambda = 532$ nm) in order to be closer to the resonance of the chain-defect modes \cite{Wake_PRL1991}. In Fig. \ref{fig03} (b) we plot the un-illuminated Raman spectra (thin line) of YBCO$_{6+x}$ for selected hole-doping levels at $T = 20$ K. For all the samples we observe multiple phonon modes between 230 and 330 cm$^{-1}$ which have been associated with Cu vibrational modes at the end of short CuO chain fragments \cite{Ivanov_PRB1995}. We note that around 600 cm$^{-1}$ (not shown) the associated oxygen vibrational modes can be observed and exhibit a similar evolution with photon dose \cite{Ivanov_PRB1995}. After full illumination (thick line) the intensity of these Raman defect modes is drastically suppressed, whereas the regular  O(2)-O(3)($z$) mode at $\sim$340 cm$^{-1}$ is not affected. In Fig. \ref{fig04} we plot the ratio of the integrated spectrum (from 200-400 cm$^{-1}$)  for the illuminated and un-illuminated cases (blue bars) as a function of hole-doping $p$. We again normalize the largest change to unity, but use negative values since the spectral intensity decreases with illumination. It is evident from these measurements that the manifestations of LIS$_1$ in the Raman spectra are most pronounced for $x \approx 0.75$ and are quickly suppressed for more underdoped samples and closer to optimal doping. We note that this doping dependence of LIS$_1$ can also be observed by using the red laser light. Since this ``bleaching'' effect was already observed at room temperature, we have not determined the onset temperatures.

\subsection{Effect of spinless impurities}

\begin{figure}[htb]
\includegraphics[width=0.925\columnwidth]{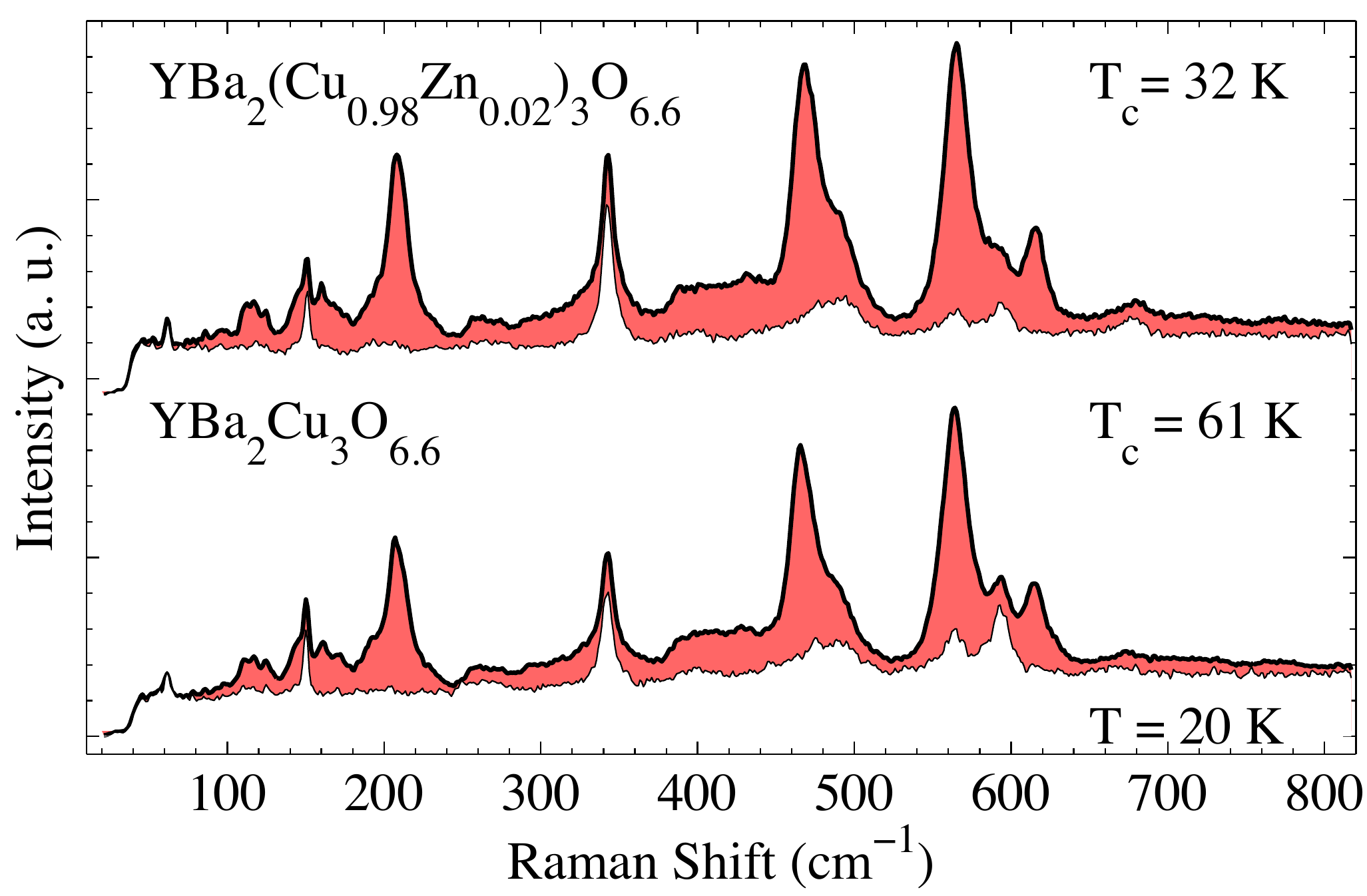}
\caption{\label{fig05}(Color online) Raman spectrum of a  2\% Zn-substituted YBCO$_{6.6}$ crystal taken with red light in the
$XX$ channel (top), compared with the spectrum of Zn-free YBCO$_{6.6}$ (bottom). Thin (thick) black lines represent un-illuminated (fully illuminated) spectra. The light-induced changes are highlighted in red. }
\end{figure}

Motivated by the observation of incommensurate magnetic order induced by the substitution of Zn$^{2+}$ impurities (which have a full $d$-electron shell and are hence nonmagnetic) for Cu$^{2+}$ in the CuO$_2$ layers, \cite{Blanco_PRL2013} we have monitored the illumination dependence of the Raman spectrum of a YBa$_2$(Cu$_{0.98}$Zn$_{0.02}$)$_3$O$_{6.6}$ crystal with $T_c = 32$ K. Neither the characteristic relaxation time nor the light-induced phonon modes exhibit any significant difference to those measured on a Zn-free sample with the same oxygen stoichiometry and $T_c = 61$ K (Fig. \ref{fig05}). 

\section{Discussion}

We start our discussion with a short summary of our main experimental findings. Our Raman measurements have revealed that a new long-lived state can be created by illuminating underdoped YBCO$_{6+x}$ ($0.43 \leq x \leq 0.93$) with laser light polarized along the crystallographic $a$-axis. This state, LIS$_2$, is characterized by the appearance of four strong new phonon features below T$_{\rm phonons} \sim 200$ K, which have been shown to resonate strongly with the red laser line~\cite{Bakr_PRB2013}. The intensities of the new phonons show a stretched-exponential dependence on illumination time (or cumulative photon dose), with a characteristic time that depends strongly on temperature. As a function of hole-doping, the intensity of the new modes is maximal around $p\!\approx 0.12$, and is drastically reduced for more lightly and more heavily doped samples.

The photon dose dependence of the LIS$_2$ and the reduction of the characteristic time $\tau$ with increasing temperature strongly suggests that it is related to light-induced reordering of O(1) oxygen atoms in the CuO chains, similar to the LIS$_1$. This assignment is further supported by the sensitivity of the LIS$_2$ to the photon polarization direction relative the CuO chain direction in the crystal structure -- the new modes in the $XX$ channel are not seen in the $YY$ channel -- and by the observation that vibrations of O(2), O(3), and O(4) atoms in or around the CuO$_2$ planes are not affected by illumination (Fig. \ref{fig03} (a)). 

Besides their origin in light-induced oxygen motion, however, the different dependences of the illumination effects on the light polarization, doping, and temperature indicate that the LIS$_1$ and LIS$_2$ are distinct from each other. In particular, the chain defect mode around 593 cm$^{-1}$ seen in the $XX$ channel (Fig. \ref{fig01} (a)), which is bleached in the LIS$_1$, remains largely unchanged in the LIS$_2$. We note that this mode is much stronger in the $YY$ channel and therefore even small changes could be monitored. However, since the  $YY$ channel measurements inevitably bleach the sample during the collection time this kind of study becomes difficult. In addition,  the LIS$_2$ is characterized by the light-induced appearance of at least four new modes in the $XX$ channel which do not appear in the LIS$_1$. The intensities of these modes and the rates of the light-induced modifications is also different in both states  (Fig. \ref{fig04}). Finally, the LIS$_2$ is only seen below $T = 200$ K, whereas the LIS$_1$ is already observed at $T = 300$ K. These differences are presumably rooted in the nature of the oxygen motion induced by the incident light field. 

While more research is required to determine the local order of the O(1) atoms in the LIS$_2$, it is likely that the CuO chain order is less pronounced than in the pristine state, and that the orthorhombicity of the crystal structure is correspondingly reduced. We note that a related light-induced shift of the ligand atoms has been observed at low temperatures in the Na$_2$[Fe(CN)$_5$NO]$\cdot$2H$_2$O molecule \cite{Haussuhl_PRL1984,Coppens_JACS1997}. 

Bakr {\it et al.} \cite{Bakr_PRB2013} had assigned the new phonons in the $XX$ channel to Brillouin-zone folding by the quasi-two-dimensional CDW, based in part on the similarity of their onset temperatures and doping dependences to the CDW order parameter determined by x-ray scattering. \cite{Blanco_PRB2014} However, as the photon dose dependence of the Raman modes was not monitored in the earlier study, the temperature dependence reported in Ref. \onlinecite{Bakr_PRB2013} reflects the uncontrolled confluence of two factors (temperature and illumination), so that its interpretation has to be reassessed. In doing so, we note that in fully illuminated samples with a saturated LIS$_2$, the intensity of the Raman modes exhibits a monotonic temperature dependence  with a sharp onset at 180-200 K and no anomaly at $T_c$ (Fig. \ref{fig01} (c)). This is at variance with the amplitude of the quasi-two-dimensional CDW measured by x-ray scattering, which shows a gradual onset at significantly lower temperatures (Fig. \ref{fig04}) as well as an intensity reduction upon cooling below $T_c$ that reflects the competition between the CDW and superconductivity\cite{Ghiringhelli_Science2012,Achkar_PRL2012,Chang_NaturePhysics2012,Blanco_PRB2014}. 

The temperature dependence of the phonon modes in fully illuminated samples (Fig. \ref{fig01} (c)) is indicative of a second-order structural phase transition. This behavior suggests an electronically driven transition to a long-range-ordered superstructure, rather than a nonequilibrium order-disorder transition of the O(1) atoms which are dilute and nearly immobile at low temperatures.  Indeed, a recent resonant x-ray scattering study of YBCO$_{6+x}$ thin films grown on SrTiO$_3$ has revealed a three-dimensionally long-range-ordered CDW state with periodicity identical to the one of the quasi-2D CDW at the same doping level \cite{Bluschke_NComm2018}. In these films, epitaxial growth disrupts the CuO chain order so that the YBCO$_{6+x}$ crystal structure is macroscopically tetragonal. The higher lattice symmetry  may enhance nesting features of the Fermi surface and stabilize CDW order (perhaps in combination with electronic reconstructions at the substrate-film interface). The 3D CDW in the YBCO$_{6+x}$ films exhibits a higher onset temperatures than its quasi-2D counterparts (still being visible at room temperature), as well as an order parameter with a monotonic temperature dependence. These features bear a striking qualitative similarity to the behavior of the phonon modes characteristic of the LIS$_2$.  We hence hypothesize that illumination with $X$-polarized photons disrupts the CuO order and induces 3D-CDW order analogous to the one observed in the films. Following the arguments presented by Bakr {\it et al.} \cite{Bakr_PRB2013}, the new phonon modes can then be understood as a consequence of Brillouin-zone folding in the CDW state. Definitive confirmation of this hypothesis will be possible in x-ray scattering experiments under light illumination \cite{note}.

The insensitivity of the light-induced phonon modes to the onset of superconductivity (Fig. \ref{fig01} (c)) and to substitution with spinless impurities (Fig. \ref{fig05}) indicates that the LIS$_2$ (and hence presumably the 3D-CDW) is not affected by the delicate competition between charge order, magnetic order, and superconductivity that shapes the phase diagram in unilluminated YBCO$_{6+x}$. \cite{Blanco_PRL2013} Further work is required to assess whether light illumination creates a phase separated state in which the 3D-CDW simply replaces superconductivity in mesoscopic or macroscopic regions, or whether the superconducting state adjusts to the presence of the more robust 3D-CDW by forming a microscopically phase-modulated state such as a pair density wave.

\section{Conclusion}

In conclusion, our Raman scattering experiments led to the  discovery of a new light-induced metastable state. The temperature and time scales of the light-induced modifications indicate that it originates in light-induced displacements of the oxygen dopants. The newly found metastable state provides a platform for a second-order structural phase transition heralded by a set of new Raman-active phonons. Comparison with x-ray scattering data on related systems suggests that the low-temperature structural phase is a three-dimensionally long-range-ordered CDW analogous to the one recently observed in bulk YBCO$_{6+x}$ crystals in high magnetic fields \cite{Gerber_Science2015} and in epitaxial YBCO$_{6+x}$ films in the absence of magnetic fields \cite{Bluschke_NComm2018}. 

If this interpretation is confirmed by complementary experiments that directly determine the CDW order parameter, illumination by red laser light polarized along the $a$-direction is a much more effective means of {\it in situ} tuning the electronic phase composition of YBCO$_{6+x}$ samples  than thermal quench-annealing \cite{Achkar_PRL2014}.  In this way, the influence of CDW order on the transport, magnetic, and optical properties of YBCO$_{6+x}$ could be systematically monitored in a wide variety of experiments (taking into account, of course, the limited light penetration depth in the samples to be investigated).  Conversely, the possible influence of light-induced oxygen reordering and CDW formation should be carefully considered in all experiments that subject YBCO$_{6+x}$ to intense visible light.

\acknowledgements{We would like to thank C.T. Lin for sample preparation 
and A. Schultz for technical assistance. }


\begin{thebibliography}{}
\bibitem{note} We also performed Raman scattering experiments on YBCO$_{6+x}$ films epitaxially grown on SrTiO$_3$, but found that the Raman response was dominated by the strong substrate contribution so that the new modes could not be conclusively identified.
\end{thebibliography}
\end{document}